\documentclass[12pt,letterpaper]{article} 
\usepackage{amssymb,amsmath}
\usepackage{graphicx}
\usepackage{enumitem}

\usepackage{multirow}
\usepackage{booktabs}
\usepackage{rotating}
\usepackage[table]{xcolor}%

\def\spacingset#1{\renewcommand{\baselinestretch}%
{#1}\small\normalsize} \spacingset{1}

\begin{document}
\spacingset{1.45}
\thispagestyle{empty}

\title{\LARGE{\bf Clustering genomic words in human DNA using
peaks and trends of distributions}}


\author{Ana Helena Tavares,
        Jakob Raymaekers,    
        Peter J. Rousseeuw,\\  
        Paula Brito,
        and Vera Afreixo
}

\date{August 14, 2018}

\maketitle

%


\bigskip		
\begin{abstract}
In this work we seek clusters of genomic words in
human DNA by studying their inter-word lag
distributions.
Due to the particularly spiked nature of these
histograms, a clustering procedure is proposed that
first decomposes each distribution into a baseline
and a peak distribution.
An outlier-robust fitting method is used to
estimate the baseline distribution (the `trend'),
and a sparse vector of detrended data captures the
peak structure.
A simulation study demonstrates the effectiveness of
the clustering procedure in grouping distributions
with similar peak behavior and/or baseline features.
The procedure is applied to investigate similarities
between the distribution patterns of genomic words
of lengths 3 and 5 in the human genome.
These experiments demonstrate the potential of the
new method for identifying words with
similar distance patterns.
\end{abstract}

\vspace{0.3cm}

\noindent
{\it Keywords:} 
Classification, 
Pattern Recognition,
Robustness,
Word distances.
\vfill

\newpage

\section{Introduction}

Genomes encode and store information that defines
any living organism. They may be represented as
sequences of symbols from the nucleotide alphabet
$\{A,C,G,T\}$.
A segment of $k$ consecutive nucleotides is
called a \emph{genomic word} of length $k$.
For each length $k$ there are $4^k$ distinct
words.

Some words have a well-defined biological
function, and several functionally important
regions of the genome can be recognized by
searching for sequence patterns, also called
`motifs' \cite{macisaac2006}.
For instance, the trinucleotide $ATG$ serves
as an initiation site in coding regions,
i.e. a marker where translation into proteins
begins~\cite{nakamoto2009}.
Also the word CG is interesting.
Although CG dinucleotides are
under-represented in the human genome,
clusters of CG dinucleotides (`CpG islands')
are used to help in the prediction and
annotation of genes~\cite{bajic2003}.
Furthermore, CpG islands are known to be
associated with the silencing of
genes~\cite{deaton2011,jacinto2007,saxonov2006}.
These examples illustrate the importance
of identifying word patterns in genomic
data.

A particular characteristic of a genomic
word is its distribution pattern.
The distribution pattern of a word along
a genomic sequence can be characterized by
the distances between the positions of the
first symbol of consecutive occurrences of
that word.
The \emph{distance distribution} of the
word is the frequency of each lag in
the DNA sequence.
Patterns in distance distributions have been
studied through several approaches (see e.g.
~\cite{afreixo2014,tavares2016,tavares2017})
and form an interesting research topic due
to their link with positive or negative
selection pressures during
evolution~\cite{burge1992,leung1996}.

In this paper we look for clusters of
genomic word distance distributions.
Because of the particularly spiked nature of
these distributions, we have developed a
3-step procedure.
First, we fit a smooth baseline distribution
using an outlier-robust fitting technique.
Secondly, we identify and characterize the
peak structure on top of that baseline.
Finally, a clustering procedure is applied
to the characterization obtained in
the first two steps.

The paper is organized as follows.
Section 2 describes distance distributions
and the proposed clustering procedure.
Section 3 is a simulation study which
measures the performance of the proposed
method.
Section 4 clusters real data, consisting of
distance distributions of words in the human
genome.
Section 5 concludes and outlines future
research directions.

\section{Methodology}
\label{sec:method}

\subsection{Word distance distributions}

In a simple random sequence with words
generated independently from an identical
distribution, the distance distribution of
a word (without overlap structure) follows
a geometric distribution~\cite{percus2002},
whose continuous approximation is an
exponential distribution.
By adding some correlation structure between
a symbol and the symbols at preceding
positions, a more refined DNA model is
obtained.
This can be achieved by assuming a $k$-th
order Markov
model~\cite{fu1996,tavares2016}.

\begin{figure}[ht]
\centering
\includegraphics[width=0.8\textwidth]
  {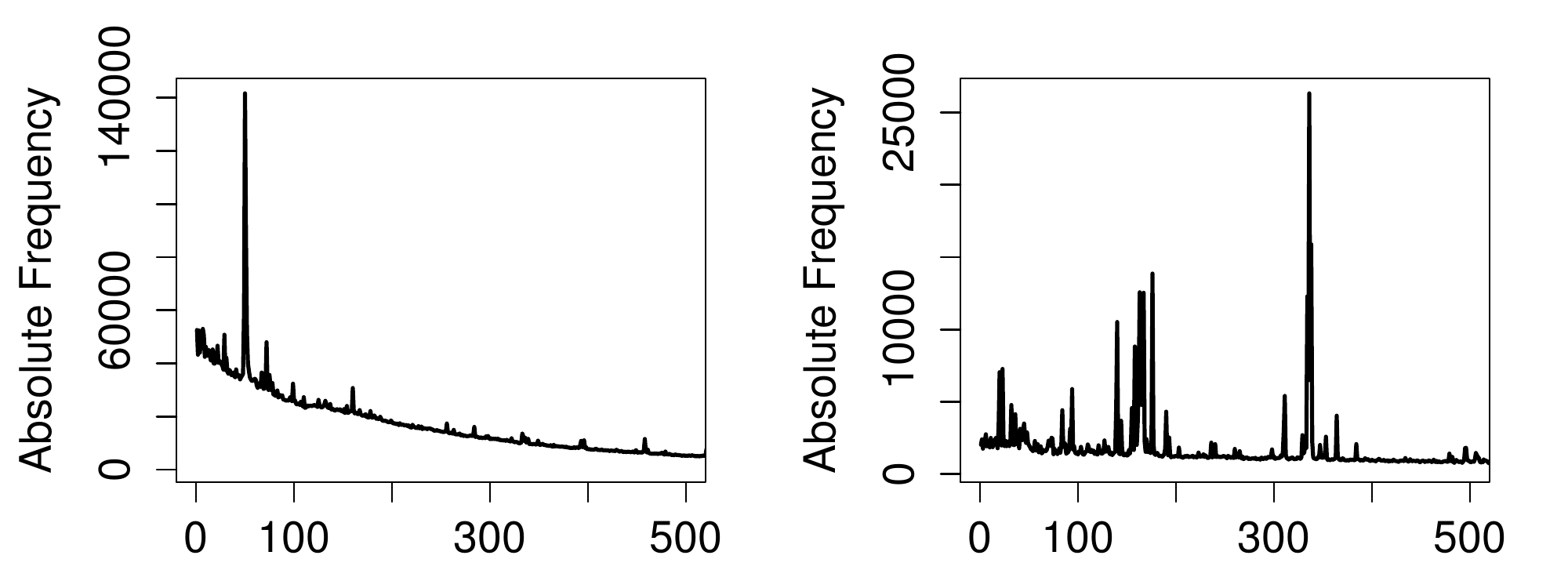}
\vskip-0.5cm
\caption{Distance distribution of the
genomic words $w=TACT$ (left) and
$w=ACGG$ (right) in the human genome.
Both distributions exhibit over-favored
distances (peaks).
The strongest peaks correspond to
distances 54 (left) and 340 (right).}
\label{fig:k4}
\end{figure}

However, real genomic sequences are
more complex and do not follow the simple
models mentioned above.
Many unexpected patterns occur in the
distance distributions of genomic words.
For instance, Figure~\ref{fig:k4} shows
the distance distributions of the words
$w=TACT$ and $w=ACGG$ in the human genome
assembly. They have strong peaks, which
correspond to distances that occur much
more often than others.

\subsection{Decomposition of distance
            distributions}
\label{sec:model}

In this study we decompose a distance
distribution into a smooth underlying
distribution (the `trend') and a peak
function.
This decomposition allows us to
separate the two essential properties
of a distribution.

Consider a genomic word $w$ of length
$k$ and denote its relative frequency
(histogram) by $f$, observed on a domain
consisting of lags
$\{k+1, k+2, \ldots, L\}$.
Note that $\sum_{j= k+1}^{L}{f(j)} = 1$.
Such a distribution typically consists of
an overall trend and some upward peaks.
Therefore, we model the distribution as
a mixture of a baseline distribution
$f_b$ and a peak function $f_{pk}$\,:
\begin{equation}
  f = f_b + f_{pk}\;\;.
\label{eq:model}
\end{equation}
We will denote the mass of the baseline
component as
$m_b=\sum_{j = k+1}^{L} f_b(j)$ and
that of the peak function as
$m_{pk}=\sum_{j = k+1}^{L} f_{pk}(j)$.
Both $f_b$ and $f_{pk}$ are nonnegative
hence $0\leq m_b\leq 1$ and
$0\leq m_{pk}\leq 1$,
with $m_b+m_{pk}=1$.

From many trial fits on distance
distributions of genomic words we
concluded that a properly scaled gamma
density function provides a good fit of
the underlying trend.
Therefore we set $f_b =\alpha f_\gamma$
with $\alpha\geq 0$ and
\begin{equation}
  f_\gamma(x; \theta, \lambda) =
	\frac{\lambda^{\theta} x^{\theta-1}
	e^{-\lambda x}}{\Gamma(\theta)}
  I(x>0)
	\label{eq:gamma}
\end{equation}
where $\theta>0$ is the shape parameter,
$\lambda>0$ is the rate parameter
(note that $1/\lambda$ is a
scale parameter), and $\Gamma(.)$ is
Euler's gamma
function~\cite{abramowitz1964}.
The gamma distribution includes the
exponential distribution as a special
case (with $\theta=1$) and can
therefore be seen as an extension of
the model in \cite{percus2002}.

The peak function $f_{pk}$ describes
the mass excess above the baseline.
If there is a peak at lag $j$ it
follows that $f_{pk}(j)=f(j)-f_b(j)$,
and if there is no peak $f_{pk}(j)=0$.

\begin{figure}[ht]
\centering
\includegraphics[width=0.8\textwidth]
  {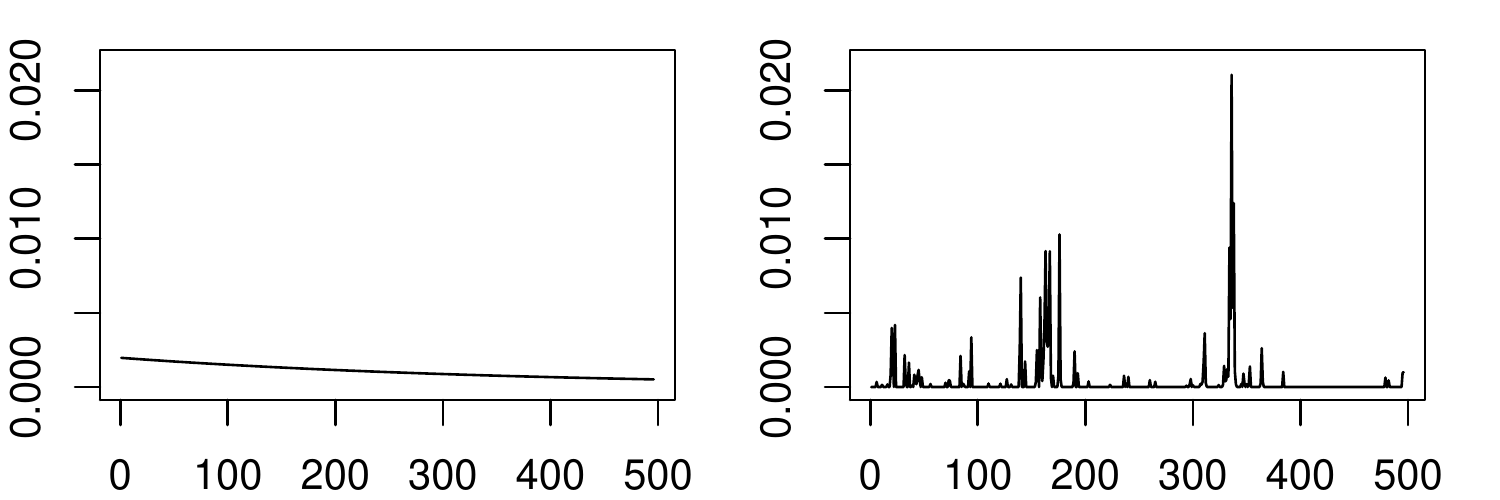}
\vskip-0.3cm
\caption{Decomposition of the distance
 distribution of the genomic word
 $w=ACGG$ into $f_b$ (left)
 and $f_{pk}$ (right).}
\label{fig:decomp1}
\end{figure}

Figure \ref{fig:decomp1} illustrates
the decomposition of the distance
distribution of the word $w = ACGG$
shown in Figure \ref{fig:k4} into a
smooth baseline function $f_b$ and a
peak function $f_{pk}$.

\subsection{Estimating the baseline}
\label{sec:LTS}

To estimate the baseline distribution
$f_b$ we need to fit a
scaled gamma curve $\alpha f_\gamma$
to the points $(j,f(j))$ of the
observed histogram, where
$j = k+1, k+2, \ldots, L$\,.
Note that $f_b$ is defined by three
parameters: $\alpha$, $\theta$ and
$\lambda$, so we have to estimate
all three together.

A first thought would be to work with
the residuals $f(j)-\widehat{f_b}(j)$,
but these suffer from
heteroskedasticity as
the variability in $f(j)$ is larger
for low $j$ than for high $j$.
In fact, if we generate $n$ data points
from the model (\ref{eq:model}) the
observed {\it absolute} frequency
$f_{ob}(j)$
at a lag $j$ in which there is no peak
follows a binomial distribution with
$n$ experiments and success
probability
$f_b(j)$\,.
(Note that in the real data $n$ is the
total number of times the word $w$
occurs in the genome.)
When the success probability is low
and $n$ is high the binomial
distribution can be well approximated
by a Poisson distribution with
mean and variance
$n\,f_b(j)$.
The standard deviation of that Poisson
distribution is thus
$\sqrt{n\,f_b(j)}$ and therefore
decreasing in $j$, which implies
heteroskedasticity of
$f_{ob}(j)-n\widehat{f_b}(j)$.
On the other hand, it is known
that the square
root of a Poisson variable has a
nearly constant standard deviation.
Therefore, we will fit the
function $\sqrt{n\,f_b}$ to the
transformed data $\sqrt{f_{ob}}$\,.
We thus use the square root as a
variance-stabilizing transform for
the Poisson distribution.
In practice, we will consider the
residuals
\begin{equation}
r(j) = \sqrt{f_{ob}(j)} -
       \sqrt{n\,\widehat{f_b}(j)}
\label{eq:res}
\end{equation}
whose standard deviation is roughly
constant at those $j$ in which
there is no peak, so we are in the
usual homoskedastic setting.

The next question is how to combine
these residuals in an objective
function to be minimized.
The standard approach for this is
the least squares (LS) objective,
which is simply the sum of all
squared residuals
$\sum_{j=k+1}^L r^2(j)$\,.
However, this does not work in our
case because of the peaks in the
data, which are outliers.
Minimizing the LS objective would
assign very high weight to the
outliers, which do not come from
the baseline $f_b$\,.
Instead we apply the least trimmed
squares (LTS) approach
of~\cite{Rousseeuw:LMS}.
This method minimizes the sum of the
$h$ smallest squared residuals, so
that
\begin{equation}
(\hat{\alpha},\hat{\theta},
 \hat{\lambda})\;\;
 \mbox{ minimizes }\;
 \sum_{i=1}^h (r^2)_{(i)}
\label{eq:LTS}
\end{equation}
where $(r^2)_{(1)} \leqslant
(r^2)_{(2)} \leqslant \ldots$
are the ordered squared residuals.
In this application we set $h$ equal
to $95\%$ of the number of values
$j$ in the domain.
By using only the $95\%$ smallest
squared residuals, the LTS method does
not fit the peaks of the distribution
and focuses only on the trend.
To avoid overemphasizing the high
lags $j$ where the fit is close to zero
and to get a more accurate fit for the
lower lags, we carry out the
LTS fit on a shorter set
$j\in\{k+1,\ldots,L^*\}$
with $L^* < L$.

\subsection{Estimating the peak
            function}
\label{sec:peak}

We now want to flag the peaks in the
observed absolute frequencies
$f_{ob}(j)$,
noting that even in lags $j$ without
a peak we do not expect $f_{ob}(j)$
to be exactly equal to
$n \widehat{f_b}(j)$ because
$f_{ob}(j)$ exhibits natural Poisson
variability with mean and variance
$n \widehat{f_b}(j)$.
Therefore we
assess the extremity of the
observed frequency $f_{ob}(j)$ by
comparing it with a high quantile
$Q(j)$ (e.g. with probability
$0.99$) of the Poisson
distribution with mean
$n \widehat{f_b}(j)$.
That is, we flag a peak at the
lag $j$ if and only if
\begin{equation}
  f_{ob}(j) > Q(j)\;\;.
\label{eq:flag}
\end{equation}
At any lag $j$ that is flagged we set
the peak function value equal to the
difference between the observed and
the expected relative frequencies,
i.e.
$f_{pk}(j) = f(j)-\widehat f_b(j)>0$.
At all the other lags we set
$f_{pk}(j) = 0$\,.

\subsection{Dimension reduction}

Suppose now that we wish to analyze
$m$ genomic words, where $m$ could be
the number of words of length $k$ in
the genome.
The raw data is then a matrix of size
$m\times (L-k)$ containing the $m$
observed lag distributions.
Each row corresponds to a discrete
distribution (a vector of length $L-k$),
denoted by $f$, which sums to one.
In the preceding subsections we
have seen how each row $f$ can be
decomposed into the sum of a baseline
and a peak function.

\begin{figure}[ht]
\centering
\includegraphics[width=0.8\textwidth]
  {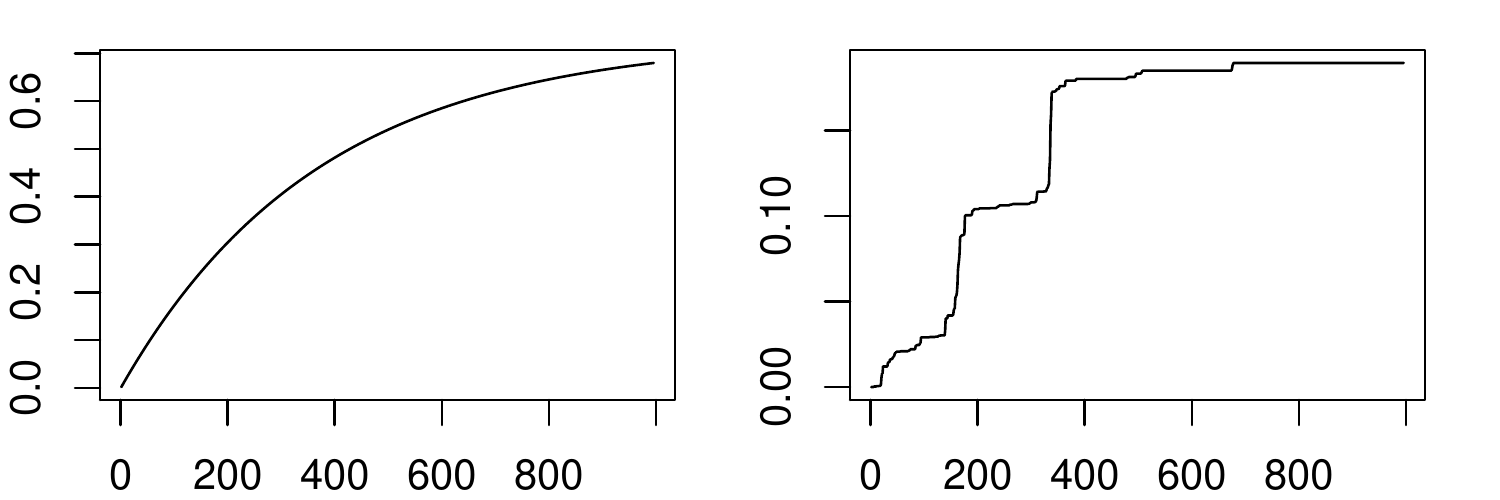}
\vskip-0.3cm
\caption{Cumulative distribution
  functions of the baseline (left)
	and the peak function (right) of
  the genomic word $w = ACGG$.}
\label{fig:decomp2}
\end{figure}

First consider the baseline functions.
In what follows we are interested in
computing a kind of distance between
such functions.
Since each baseline function $f_b$
is characterized by a triplet of
parameters $(\alpha,\theta,\lambda)$,
a simple idea would be to compute
the Euclidean distance between such
triplets.
However, the three parameters have
different scales, and triplets with
relatively high Euclidean distance
can describe similar-looking curves
and vice versa.
To remedy this, we first construct
the cumulative distribution function
(CDF) of each baseline, given by
$F_b(j) = \sum_{i=k+1}^{j} f_b(i)$
for $j = k+1, \ldots, L$.
The left panel of Figure
\ref{fig:decomp2} illustrates this
for the word $w = ACGG$, the lag
distribution of which was shown
in Figure \ref{fig:k4} and
decomposed in Figure
\ref{fig:decomp1}.
Note that $F_b(L) = m_b < 1$ when
$m_{pk} > 0$.

We can then think of the Euclidean
distance between two CDFs $F_b$ and
$G_b$ as a way to measure their
dissimilarity.
Note that these CDFs still have
$L-k$ dimensions, which is
usually very high.
Therefore, in the second step we
apply a principal component analysis
(PCA) to these $m$ high-dimensional
vectors.
This operation preserves much of
the Euclidean distances.
The number of components we retain,
$q_b$\,, is selected such that at
least a given percentage of the
variance is explained.
Typically $q_b \ll L-k$ so the
dimension is reduced substantially.
The scores associated to the first
$q_b$ components yield a data
matrix of
much smaller size $m\times q_b$\,.
Note that these scores are
uncorrelated with each other
by construction.

For the peak functions, stacking the
$m$ rows on top of each other also
yields a matrix of size
$m\times (L-k)$.
This data matrix is sparse in the
sense that few of its elements are
nonzero.
We then follow the same strategy to
that used for the baseline functions:
first we convert the peak functions
to CDFs as illustrated in the right
panel of Figure \ref{fig:decomp2},
and then we apply PCA
yielding $q_{pk}$ components, where
$q_{pk}$ is selected to attain at
least a given explained variance.
The resulting score matrix has size
$m\times q_{pk}$\,.

\subsection{Clustering}
\label{sec:clusproc}

Clustering, also known as unsupervised
classification, aims to find groups in
a dataset (see e.g.~\cite{kaufman1990}).
Here our dataset is a matrix of size
$m \times (q_b + q_p)$ obtained by
applying the above preprocessing to all
of the $m$ frequency distributions.
We explore clustering based on only
the peak component \emph{(Method 1)},
only the baseline component
\emph{(Method 2)}, and based on both
\emph{(Method 3)}.
To each of these datasets we apply
the k-means method, in which $k$
stands for the number of clusters
which is specified in advance.
(The letter `k' in the name of this
method differs from the word length
$k$ used elsewhere in this paper.)
This approach defines the center of
a cluster as its mean, and assigns each
object to the cluster with the nearest
center. Its goal is to find a partition
such that the sum of squared distances
of all objects to their center is as
small as possible.
The algorithm starts from a random
initialization of cluster centers and
then iterates from there to a local
minimum of the objective function.
This is not necessarily the global
minimum.
As a remedy for this problem, multiple
initial configurations are generated
and iterations are applied to them,
after which the final solution with
the lowest objective is retained.

Since k-means looks for spherical
clusters, it works best when the input
variables are uncorrelated and have
similar scales. The preprocessing by
PCA in the previous step has created
uncorrelated variables, and in our
experiments their scales were of the
same order of magnitude.

\subsection{Selecting the number
            of clusters}

The result of $k$-means clustering
depends on the number of clusters $k$,
which is often hard to choose a priori.
Therefore it is common practice to
run the method for several values of
$k$, and then select the `best' value
of $k$ as the one which optimizes a
certain criterion called a
validity index.
Many such indices have been proposed
in the literature.
Here we will focus on three of them:
the Calinski-Harabasz (CH) index, the
C index, and the silhouette (S) index.

The CH index~\cite{calinski1974}
evaluates the clustering based on the
average between- and within-cluster
sums of squares. The approach selects
the number of clusters with the highest
CH index.

The C index reviewed
in~\cite{hubert1976} relates the sum of
distances over all pairs of points from
the same cluster (say there are $N$ such
pairs) to the sum of the $N$ smallest and
the sum of the $N$ largest distances
between pairs of points in the entire data
set. It ranges from 0 to 1 and should be
minimized. To compute the C index all
pairwise distances have to be computed and
stored, which can make this index
prohibitive for large datasets.

The S index~\cite{rousseeuw1987} is the
average
silhouette width over all points in the
dataset. The silhouette width of a point
relates its average distance to points
of its own cluster to the average distance
to points in the `neighboring' cluster.
The silhouette index ranges from $-1$ to
$+1$ and large values indicate
a good clustering.

The performance of these measures
depends on various data
characteristics. An early reference for
comparing clustering indices is
\cite{milligan1985examination}, which
concludes that CH and C exhibit
excellent recovery characteristics in
clean data (the S index was not yet
proposed at that time).
More recent works evaluate clustering
indices also in datasets with outliers
and noise, see e.g.
\cite{guerra2012,liu2010}).
Guerra et al.~\cite{guerra2012} rank
CH and S in top positions, and
report poor performance of the C
index in that situation.

Rather than choosing one of these
indices we will compute all three in
our study, and plot each of them
against the number of clusters.
The local extrema in these curves
can be quite informative.


\section{Simulation study}

To better understand the behavior
of the proposed procedure, a
simulation study is performed.
To assess how well a clustering
method performs, we compute a
measure of agreement between the
resulting partition and the
true one.

\subsection{Study design}

Experiments are performed on
datasets consisting of three
distinct groups of discrete
distributions, denoted by $G_1$,
$G_2$ and $G_3$, whose
characteristics are defined by a
five factor factorial design.
The factors and levels used in
the study are listed
in Table~\ref{tab:design}.
They have the following meaning.
\begin{itemize}
\item Trend ($T$) is defined by
  the Gamma parameters $\theta$
	(shape) and $\lambda$ (rate).
	When $T$ is `same' the
	distributions in all groups
	have the same baseline
  parameters.
\item Number of peaks ($NP$) gives
  the number of peaks generated
	in each distribution.
	When $NP$ is `same' all
	distributions exhibit the same
	number of peaks, $np$, set as 10.
	In case $NP$ is `distinct'
	the number of peaks
	is set to 20 in $G_1$, 10 in
	$G_2$, and 5 in $G_3$.
\item Peak locations ($PL$). In
  each group the `mean locations'
  ($ml$) are generated uniformly
	on the domain. For each
	member of that group the peak
  locations are generated around
	the mean locations of that group
	($ml\pm h$, with $h=3$).
	When $PL$ is `similar'
	all groups have the same mean
	locations.
\item Peak mass ($PM$) corresponds
  to the amount of mass $m_p$ in
	the peaks of the distribution,
	so the mass of the baseline is
	$1-m_p$. Three levels are
  considered: distributions of all
	groups have the same $m_p>0$;
	distributions of distinct groups
	have different $m_p>0$;
	distributions of $G_1$ and $G_2$
	have different $m_p>0$ and
	distributions from $G_3$ have
	$m_p=0$.
	Note that the factors $NP$ and
	$PM$ are not independent, as
	$NP=1$ implies $PM\neq3$, and
	$PM=3$ implies that the
	distributions in $G3$ have no
	peaks ($np=0$).
\item Sample size ($SS$) describes
  the number of elements in each
	group. In the `balanced' setting
	all groups have the same number
	of distributions.
\end{itemize}

Each simulated distribution is
constructed from a baseline
function and a peak function.
All distributions belonging to the
same group have the same factor
levels.

\begin{table}[htbp]
\centering \caption{Factors of
	the experimental study and
	corresponding levels.
  Factors: trend, $T$; number of
	peaks, $NP$; peak locations,
	$PL$; peak mass, $PM$;
  sample size per group, $SS$.}
\renewcommand{\arraystretch}{1.0}
\setlength\tabcolsep{2.5pt}
\normalsize
\vskip0.3cm
    \begin{tabular}{llrrrr}
    \toprule
    \multirow{2}[2]{*}{Factor} & \multicolumn{1}{l}{\multirow{2}[2]{*}{Level}} & \multicolumn{1}{c}{Para-} & \multicolumn{3}{c}{Groups} \\
           & \multicolumn{1}{c}{} & \multicolumn{1}{c}{meters} & G1     & G2     & G3 \\
           \midrule
    \multirow{4}[2]{*}{Trend (T)} & \multicolumn{1}{l}{\multirow{2}[1]{*}{1. same}} & $\theta$ & 0.8    & 0.8    & 0.8 \\
           & \multicolumn{1}{c}{} & $\lambda$ & 0.0005 & 0.0005 & 0.0005 \\
           & \multicolumn{1}{l}{\multirow{2}[1]{*}{2. distinct}} & $\theta$ & 0.6    & 0.8    & 0.95 \\
           & \multicolumn{1}{c}{} & $\lambda$ & 0.0001 & 0.0005 & 0.001 \\
    \hline
    Number of Peaks  & 1. same & $np$ & 10     & 10     & 10$^\ast$ \\
    ($NP$)             & 2. distinct & $np$ & 20     & 10     & 5$^\ast$ \\
    \hline
    Peak Locations & 1. similar & -      & -      & -      & - \\
    ($PL$)         & 2. distinct & -      & -      & -      & - \\
    \hline
    \multirow{3}[2]{*}{Peak Mass (PM)} & 1. same & $m_p$ & 0.05   & 0.05   & 0.05 \\
           & 2. distinct & $m_p$ & 0.1    & 0.05   & 0.02 \\
           & 3. distinct with 0 & $m_p$ & 0.1    & 0.05   & 0 \\
    \hline
    \multirow{2}[2]{*}{Sample Size (SS)} & 1. balanced &       & 200    & 200    & 200 \\
           & 2. not balanced &       & 50    & 150    & 400 \\
    \bottomrule
    \end{tabular}\\
\begin{flushleft}
$^\ast$These values are replaced by 0 in case factor $PM$ takes level 3.
\end{flushleft}
\label{tab:design}%
\end{table}%

Note that for the baseline function
(\ref{eq:gamma}) only the parameters
$\theta$ and $\lambda$ are
user-defined, while $\alpha$ is not.
This is because $\alpha$ is
determined from the peak mass
$m_p$ by
\begin{equation}
  \alpha=(1-m_p)/\sum_{j=k+1}^L
	f_\gamma(j;\theta,\lambda)\;\;.
\label{eq:alpha}
\end{equation}
Therefore the baseline functions are
determined by the trend $T$ and the
total peak mass $PM$.
Since the baseline construction
depends on $PM$, it is required that
the peak mass takes the same value in
all groups ($PM$=1) in order to
obtain similar baselines ($T$=1).\\

We will say that groups have
\emph{similar baselines} when their
$T$ is `same' and peak mass $PM$
is `same', and that they have
\emph{distinct baselines} when
$T$ is `distinct'.
Also, when number of peaks $NP$
is `same' and peak location $PL$
is `similar', we will say that
the groups have \emph{similar peak
functions}, and when $PL$ is
`distinct' they are said to have
\emph{distinct peak functions}.\\

We are interested in the following
three scenarios:
\begin{itemize}
\item[] \textbf{Scenario 1} -
  Groups have similar baselines
	and distinct peak functions;
\item[] \textbf{Scenario 2} -
  Groups have similar peak functions
	and distinct baselines;
\item[] \textbf{Scenario 3} -
  Groups have distinct baselines
  and distinct peak functions.
\end{itemize}
The remaining case where both the
baselines and the peaks are similar
is not of interest since its groups
are basically the same.

The combination of the three scenarios
of interest with the possible levels
of the design factors leads to 20
possible data configurations: 4 cases
for scenario 1, 4 cases for scenario 2
and 12 cases for scenario 3, as can be
seen in Table~\ref{tab:scenarios}.
For each case 100 independent samples
were generated, and the clustering
methods described in
section~\ref{sec:clusproc} were
applied to each sample.

\begin{table}[htb]
\centering
\caption{Possible combinations of
  factor levels, leading to 20 data
  conditions, \mbox{organized} by 
	scenario (1, 2 or 3).}
\vskip0.7cm
\renewcommand{\arraystretch}{1.0}	
\setlength\tabcolsep{2.5pt}
\normalsize
\begin{tabular}{ccp{2cm}p{4.5cm}p{4.5cm}}		
    \cline{4-5}
           &        &        & \multicolumn{2}{c}{ Peak Functions} \\
           &        &        & \multicolumn{1}{c}{Similar } & \multicolumn{1}{c}{Distinct} \\
    \cline{3-5}
           &        & \multicolumn{1}{c}{Factor}       & \multirow{2}{*}{$\qquad NP$=$PL$=1} & \multirow{2}{*}{$\qquad PL\neq$1} \\
           &        & \multicolumn{1}{c}{Levels}       &                  &  \\
    \cline{1-5}
    \multicolumn{1}{c}{\multirow{8}[4]{*}{\begin{sideways}Baselines\end{sideways}}} & \multicolumn{1}{c}{\multirow{4}[2]{*}{\begin{sideways}Similar\end{sideways}}} &     &   \cellcolor{gray!25}  & \multicolumn{1}{c}{\textbf{Scenario 1}} \\
    \multicolumn{1}{c}{} & \multicolumn{1}{c}{} & \multicolumn{1}{c}{$T=1$} & \cellcolor{gray!25}      & $T=1$; $NP\in\{1,2\}$; \\
    \multicolumn{1}{c}{} & \multicolumn{1}{c}{} & \multicolumn{1}{c}{and} & \cellcolor{gray!25}        & $PL=2$; $PM=1$;\\
    \multicolumn{1}{c}{} & \multicolumn{1}{c}{} & \multicolumn{1}{c}{$PM=1$} & \cellcolor{gray!25}     & $SS\in\{1,2\}$\\
    \cline{3-5}
    \multicolumn{1}{c}{} & \multicolumn{1}{c}{\multirow{4}[2]{*}{\begin{sideways}Distinct\end{sideways}}} &  &\multicolumn{1}{c}{\textbf{Scenario 2}}    & \multicolumn{1}{c}{\textbf{Scenario 3}}\\
    \multicolumn{1}{c}{} & \multicolumn{1}{c}{} & \multicolumn{1}{c}{}      & $T=2$; $NP=1$;           & $T=2$; $NP\in\{1,2\}$;  \\
    \multicolumn{1}{c}{} & \multicolumn{1}{c}{} & \multicolumn{1}{c}{$T=2$} & $PL=1$; $PM\in\{1,2\}$;  & $PL=2$; $PM\in\{1,2,3\}$; \\
    \multicolumn{1}{c}{} & \multicolumn{1}{c}{} & \multicolumn{1}{c}{}      & $SS\in\{1,2\}$           & $SS\in\{1,2\}$\\
\bottomrule
\end{tabular}
\vskip0.5cm
\label{tab:scenarios}
\end{table}

\subsection{Data generation}
The data sets were generated according
to the corresponding levels of the
factors $T$, $NP$, $PL$, $PM$ and $SS$.
All data sets consist of $m=600$
discrete distributions on $L=1500$
lags, with their peaks located in the
first 1000 lags.
The distributions are labeled by group
($G_1$, $G_2$ and $G_3$).

{\it Baseline distribution.}
The baseline distributions $f_b$ are
given by $\alpha$ times the gamma
density $f_\gamma(\theta,\lambda)$
of (\ref{eq:gamma}).
The gamma parameters $\theta$ and
$\lambda$ are determined by the
factor $T$ with parameter values
shown in Table \ref{tab:design},
plus Gaussian noise. The formula is
\begin{equation}
f_b(j) = \alpha f_\gamma(j;
         \theta+\delta_\theta,
				 \lambda+\delta_\lambda)
\end{equation}
where $\delta_\theta\sim N(0,0.01)$,
$\delta_\lambda\sim N(0,0.00001)$
and $\alpha$ is
determined from the triplet
$(\theta+\delta_\theta,
  \lambda+\delta_\lambda, m_p)$
according to (\ref{eq:alpha}).\\

{\it Peak function.}
To define a peak function $f_{pk}$
we first determine the peak locations
from the factors $PL$ and $NP$
(as described above), and their
magnitudes from $PM$ and $T$.
In all non-peak positions the
peak function is set to zero.\\

{\it Sampling variability.}
The generated baseline function and
peak function together yield a
discrete distribution $f$ as in
formula (\ref{eq:model}).
We then sample a dataset with 50,000
observations from this population
distribution, in a natural way.
We first construct the CDF of $f$,
given by
$F(j) = \sum_{i \leqslant j} f(i)$
for all $j$ in the domain.
Then we consider the quantile
function denoted as $F^{-1}$:
for each value $u$ in $]0,1[$ we set
$F^{-1}(u)=\min\{ j\, ;\,
 F(j) \geqslant u\}$.
This quantile function takes only a
finite number of values.
Now we draw 50,000 random values
from the uniform distribution on
$]0,1[$ and apply $F^{-1}$ to each,
which yields 50,000 lags in the
domain that are a random sample
from the distribution $f$ given by
~(\ref{eq:model}).
This sample forms an empirical
probability function $f_{ob}$\,.
We then apply the procedure of
Section \ref{sec:method} to carry
out a clustering on 600 such
empirical distributions.

\subsection{Performance evaluation}

Each replication takes a set of 600
distributions and returns a partition
of these data.
To assess the performance of the
method, a measure of agreement between
the resulting partition and the
true partition is needed.
Milligan and
Cooper~\cite{milligan1986study}
evaluated different indices for
measuring the agreement between
partitions and recommended the
Adjusted Rand Index (ARI), introduced
in~\cite{hubert1985}.
The ARI takes values between -1 and 1,
has a maximum value of 1 for matching
classifications and has an expected
value of zero for random
classifications. For each case we
report the mean and standard deviation
of ARI over the 100 replications.

\subsection{Results}

Table \ref{tab:ARI} summarizes the
results of the simulation.
Each row in the table corresponds to
a particular case, determined by the
levels of the 5 factors (T, NP,
PL, PM, SS). The rows are grouped
by the 3 scenarios listed in
Table \ref{tab:scenarios}.
Scenario 1 has distinct peak
functions, scenario 2 has distinct
baselines, and scenario 3 has both.

The first columns of
Table \ref{tab:ARI} describe the
factor levels, followed by columns
for each of the three methods.
In each of those the mean and the
standard deviation (in parentheses)
of the Adjusted Rand Index over the
100 replications are listed.
The final columns  list the number of
principal components retained for the
baselines (b) and the peak functions
(pk). These numbers were obtained by
requiring that the percentage of
explained variance is at least 90\%.
We see that the baselines require
only 2 components. For the peaks the
number is high when the peak masses
are the same (PM=1) and low otherwise
(in the latter case it requires
few PCs to explain the larger
peaks).

\begin{table}[!ht]
\centering
\caption{Mean and standard deviation
of the Adjusted Rand Index obtained
from 100 replicas of each case.
Results are organized by scenario and
method.
Each case is defined by a
combination of five factors: trend, T;
number of peaks, NP; peak locations,
PL; peak mass, PM; and sample size
per group, SS.
The final columns list the number of
principal components retained for the
baselines (b) and the peak functions
(pk).}
\vskip0.5cm
\renewcommand{\arraystretch}{0.8}
\setlength\tabcolsep{3pt}
\normalsize
    \begin{tabular}{rrrrrrrrrrrrrrrrr}
    \toprule
    \multicolumn{5}{c}{\textbf{Factors}}         &  & \multicolumn{2}{c}{\textbf{Method 1}} &   & \multicolumn{2}{c}{\textbf{Method 2}} &  & \multicolumn{2}{c}{\textbf{Method 3}} &  & \multicolumn{2}{c}{\#\textbf{PC}} \\
    \multicolumn{1}{l}{\small{T}} & \multicolumn{1}{l}{\small{NP}} & \multicolumn{1}{l}{\small{PL}} & \multicolumn{1}{l}{\small{PM}} & \multicolumn{1}{l}{\small{SS}}  & &  && &     && &     && & \multicolumn{1}{r}{b} & \multicolumn{1}{r}{pk}\\

    \midrule
    \multicolumn{5}{l}{Scenario 1:}             &&    \\
    1      & 1      & 2      & 1      & 1      && 0.989  & (0.046) &        & 0.000  & (0.003) &        & 0.817  & (0.255) && 2 & 62 \\
    1      & 1      & 2      & 1      & 2      && 0.886  & (0.224) &        & 0.000  & (0.007) &        & 0.493  & (0.245) && 2 & 58 \\
    1      & 2      & 2      & 1      & 1      && 0.987  & (0.052) &        & 0.000  & (0.002) &        & 0.837  & (0.245) && 2 & 55 \\
    1      & 2      & 2      & 1      & 2      && 0.821  & (0.245) &        & -0.002 & (0.007) &        & 0.530  & (0.208) && 2 & 39 \\
    \multicolumn{5}{l}{Scenario 2:}             && \\
    2      & 1      & 1      & 1      & 1      && 0.082  & (0.131) &        & 0.966	& (0.019) &        & 0.969  & (0.018)  && 2 & 42 \\
    2      & 1      & 1      & 1      & 2      && 0.043  & (0.085) &        & 0.934	& (0.036) &        & 0.940  & (0.036)  && 2 & 45 \\
    2      & 1      & 1      & 2      & 1      && 1.000  & (0.000) &        & 0.987	& (0.008) &        & 1.000  & (0.000)  && 2 & 3 \\
    2      & 1      & 1      & 2      & 2      && 1.000  & (0.000) &        & 0.989	& (0.008) &        & 1.000  & (0.000)  &&  2 & 2 \\
    \multicolumn{5}{l}{Scenario 3:}             &&     \\
    2      & 1      & 2      & 1      & 1      && 0.976  & (0.060) &        & 0.965  & (0.019) &        & 0.999  & (0.002) && 2 & 58 \\
    2      & 1      & 2      & 1      & 2      && 0.919  & (0.183) &        & 0.988  & (0.009) &        & 1.000  & (0.000) && 2 & 58 \\
    2      & 1      & 2      & 2      & 1      && 1.000  & (0.000) &        & 0.992  & (0.006) &        & 1.000  & (0.000) && 2 & 5 \\
    2      & 1      & 2      & 2      & 2      && 1.000  & (0.000) &        & 0.998  & (0.003) &        & 1.000  & (0.000) && 2 & 5 \\
    2      & 1      & 2      & 3      & 1      && 1.000  & (0.000) &        & 0.933  & (0.036) &        & 0.999  & (0.004) && 2 & 3 \\
    2      & 1      & 2      & 3      & 2      && 1.000  & (0.000) &        & 0.988  & (0.008) &        & 1.000  & (0.000) && 2 & 2 \\
    2      & 2      & 2      & 1      & 1      && 0.989  & (0.030) &        & 0.989  & (0.008) &        & 1.000  & (0.000) && 2 & 56 \\
    2      & 2      & 2      & 1      & 2      && 0.900  & (0.203) &        & 0.992  & (0.007) &        & 1.000  & (0.000) && 2 & 40 \\
    2      & 2      & 2      & 2      & 1      && 1.000  & (0.000) &        & 0.997  & (0.004) &        & 1.000  & (0.000) && 2 & 6 \\
    2      & 2      & 2      & 2      & 2      && 1.000  & (0.000) &        & 0.964  & (0.022) &        & 0.999  & (0.002) && 2 & 4 \\
    2      & 2      & 2      & 3      & 1      && 1.000  & (0.000) &        & 0.929  & (0.042) &        & 0.999  & (0.002) && 2 & 3 \\
    2      & 2      & 2      & 3      & 2      && 1.000  & (0.000) &        & 0.988  & (0.009) &        & 1.000  & (0.000) && 2 & 2 \\
\bottomrule
\end{tabular}
\label{tab:ARI}
\end{table}

\paragraph{Method 1}
The first method applies the clustering
to the PCA scores obtained from the
peak functions. Therefore, good
performance is expected in scenarios
with distinct peak locations between
the groups (scenarios 1 and 3).
Indeed, Method 1 performs very well
in scenario 1
($\mbox{ARI}\geqslant 0.821)$ and
scenario 3
($\mbox{ARI}\geqslant 0.900$).

In scenario 2 the peak locations
are the same. In the first two cases
the peak masses are similar and in
the other two cases the peak masses
are distinct. As expected, Method 1
recovers the peak differences in the
latter cases, whereas there are no
differences to recover in the former.

\paragraph{Method 2}
This method clusters the PCA scores
of the baselines, so it is expected
to work well in scenarios 2 and 3 in
which the trends are distinct, and
not in scenario 1 in which the
baselines are similar. The simulation
results confirm this, as the groups
are not recovered in scenario 1
($\mbox{ARI}\approx 0$) and are
identified with high accuracy in
scenarios 2 and 3
($\mbox{ARI}\geqslant 0.929$).

\paragraph{Method 3}
The input for Method 3 are the
scores of the baselines as well as
those of the peaks, and indeed it
is the best performer in scenario
3 where the groups have distinct
baselines combined with distinct
peaks
($\mbox{ARI}\geqslant 0.999$).
In that scenario it is also good
at distinguishing groups with
peaks from groups without peaks
($PM=3$).
Also in scenario 2 we see that
Method 3 works well, in fact it
even slightly outperforms the other
methods in that situation.
Only in scenario 1 does Method 3
perform less well. It is still
fine when the groups have balanced
sizes ($SS=1$) but becomes weaker
when the groups are unbalanced
($SS=2$).

\begin{figure}[ht]
\centering
\includegraphics[width=0.75\textwidth]
  {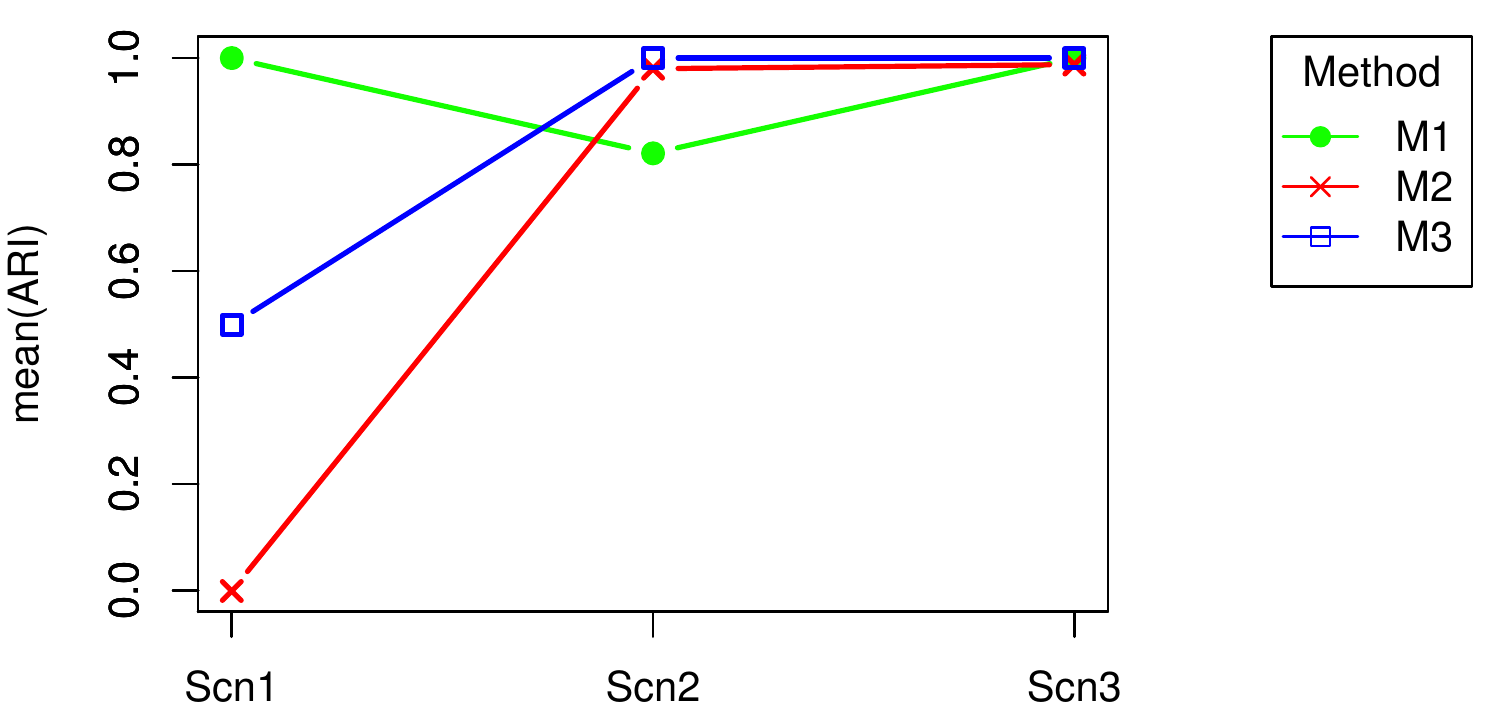}
\vskip-0.4cm
\caption{Performance of each method,
given by the mean ARI of all replications
from cases in each scenario: groups
with similar trends and distinct peak
locations (scenario 1); groups with
distinct trends and similar peak
locations (scenario 2); and groups
with distinct trends and distinct peak
locations (scenario 3). The clustering
methods 1, 2 and 3 correspond to the
three broken lines.}
\label{fig:ARI}
\end{figure}

Figure~\ref{fig:ARI} provides
a rough summary of the simulation
results by showing the ARI
averaged over all cases of each
scenario.
The performance of a method is
thus measured by three numbers.
We note that no method is best
in all scenarios.
Method 2, which ignores the peak
information, is never the best
method.
Method 1 is the best in scenario
1, and Method 3 is the best in
scenarios 2 and 3.
For a given dataset it is
recommended to carry out a
preliminary inspection to
determine which scenario it
corresponds to, before selecting
the clustering method.


\section{Application to real data}

In this section we analyze two
datasets, consisting of the lag
distributions of all words of length
$k = 3$ and $k = 5$ in the complete
human genome. These datasets are
denoted by $DD_k$ where $k$
identifies the word length.
$DD_3$ consists of 64 distributions
and $DD_5$ contains 1024
distributions.
A preliminary visual inspection of
these histograms revealed that
there are substantial differences
in both the trends and the peak
structures, so in accordance with
the conclusions of the simulation
study
we selected Method 3 (described in
Section~\ref{sec:clusproc}) for
clustering the words in each
dataset.

\subsection{Data and data processing}

We used the complete DNA sequence
of the human genome assembly,
downloaded from the website of the
National Center for Biotechnology
Information. The available assembled
chromosomes (in version GRCh38.p2)
were processed as separate sequences
and all non-ACGT symbols were
considered as sequence separators.

The counts of word lags were
obtained by a dedicated C program
able to handle large datasets
(the haploid human genome has
over 3 billion symbols).
We analyzed the absolute
frequences of the lags
$j=k+1,\ldots,L$ where $L=1000$
for $k=3$ and $L=4000$ for $k=5$.

The R language was used to decompose
the lag distributions, to perform
the principal component analysis
and the clustering and to carry out
further statistical analysis.
The R code used in this report, as 
well as the data sets and a script 
analyzing them and reproducing the 
figures can be downloaded from\\
\mbox{{\it https://wis.kuleuven.be/stat/robust/software}}\,.

\subsection{Decomposing the lag
  distributions}

In both datasets we first estimated
the baseline distribution by LTS
as described in Subsection
\ref{sec:LTS}, in which we set
$L^*=200$ for $DD_3$ and
$L^*=1500$ for $DD_5$.
The peak functions were then estimated
as described in Subsection
\ref{sec:peak}.

\subsection{Clustering words of
            length 3}

Each distribution in $DD_3$ is
summarized by 4 values, as the
PCA retains 2 components for the
peaks and 2 components for the
baselines.

\begin{figure}[ht]
\centering
\includegraphics[width=1.0\textwidth]
  {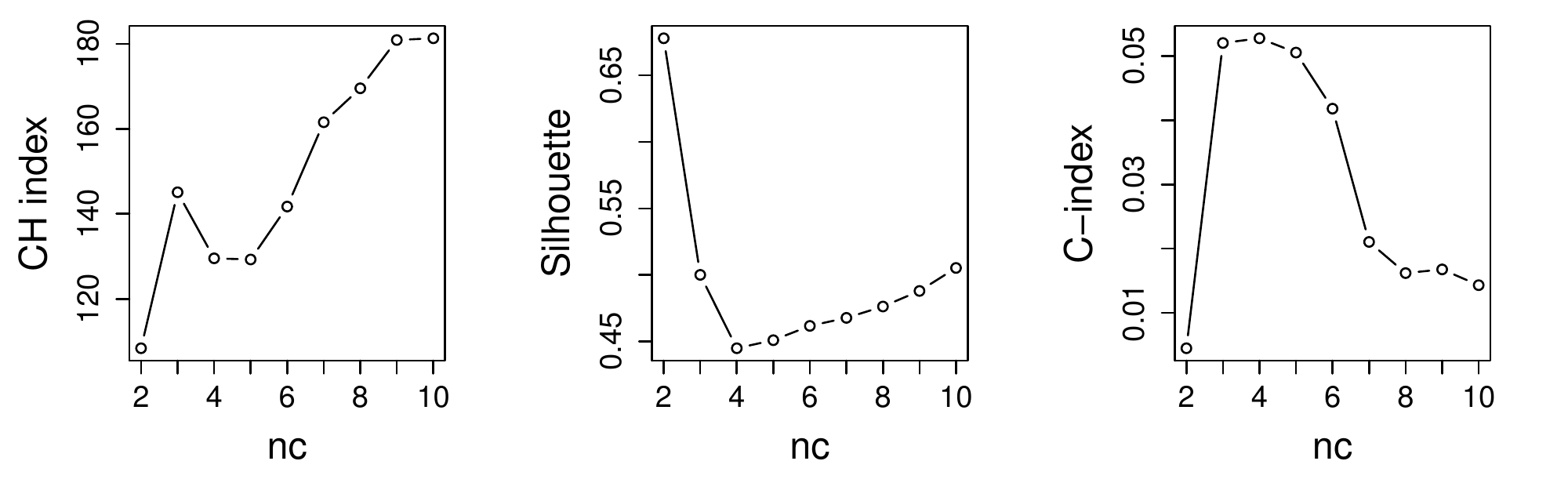}
\vskip-0.7cm
\caption{Validation indices for
clustering $DD_3$ by the number of
clusters $nc$: the Calinsky-Harabasz
index (left), silhouette coefficient
(center) and C-index (right). }
\label{fig:indexk3_10}
\end{figure}

Figure \ref{fig:indexk3_10} plots
the validation indices against
the number of clusters ($<10$).
The CH index has a local maximum
at 3 clusters and is high again
at 6 clusters or more, whereas
the silhouette index is highest
for 2 clusters and the C index
is lowest (best) for 2 clusters
and gets low again for over 6
clusters. From the 3 indices
together it would appear natural
to select 2 clusters, for which
$\mbox{CH}=108$, $S=0.68$  and
$C=0.052$.
The cluster $C_1$ has 8 elements,
and cluster $C_2$ has 56.

To test the stability of this
clustering we follow the approach
of Hennig \cite{hennig2008}.
We draw a so-called bootstrap
sample, which is a random sample
with replacement from the 64
objects in the $DD_3$ dataset.
This creates a different dataset
with 64 objects, some of which
coincide. We then
apply the same clustering method
to it, set to 2 clusters. Let us
call the new clusters $D_a$ and
$D_b$. Then we compute the
so-called Jaccard similarity
coefficient of $C_1$ with the
new clustering, defined as
\begin{equation}
  J(C_1) = \mbox{max} \Big(
	\frac{|C_1 \cap D_a|}
	     {|C_1 \cup D_a|}\,,
	\frac{|C_1 \cap D_b|}
	     {|C_1 \cup D_b|}\Big)
	\leqslant 1			
\label{eq:Jaccard}
\end{equation}
where $|\ldots|$ stands for the
number of elements.
A high value $J(C_1)$ indicates
that $C_1$ is similar to one of
the clusters of the new partition.
We compute $J(C_2)$ analogously.
Then we repeat this whole
procedure for a new bootstrap
sample and so on, 200 times in all.
The average of the 200 values of
$J(C_1)$ equals 0.952,
which means that the cluster $C_1$
is very stable.
For cluster $C_2$ we attain the
stability index $0.978$
which is even higher.
\begin{figure}[htb]
\centering
\includegraphics[width=0.9\textwidth]
  {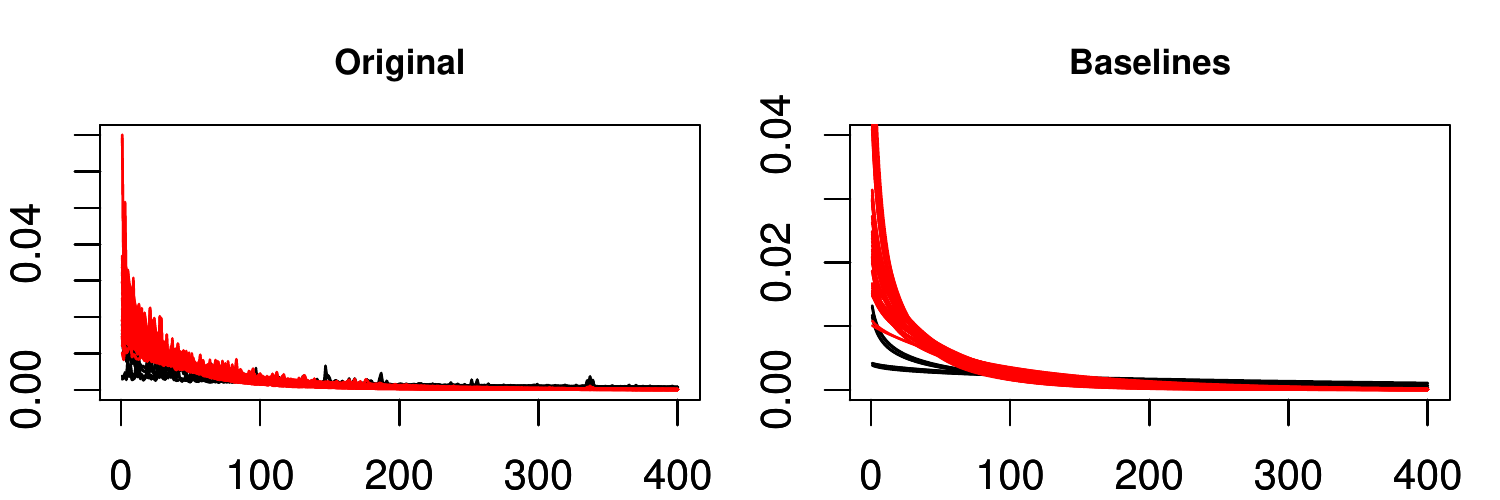}
\vskip-0.3cm
\caption{Clustering of the dataset
$DD_3$ in clusters $C_1$ and $C_2$.
The lag distributions are shown on
the left, and the corresponding
baselines on the right. Cluster
$C_1$ is in black and $C2$ in red.}
\label{fig:clustk3_2}
\end{figure}

Figure \ref{fig:clustk3_2} depicts
the clusters $C_1$ and $C_2$.
The lag distributions in $C_1$ are
flatter than those in $C_2$.
It turns out that all the words in
$C_1$ contain the dinucleotide CG
(known as CpG).
In fact, $C_1$ consists exactly of
the 8 words of length 3 that
contain CG (i.e., ACG, CCG, GCG,
TCG, CGA, CGC, CGG, CGT), so
$C_2$ contains no words with CG.
The special behaviour of the CG
dinucleotide in the human genome is
well reported in the literature.
Although human DNA is generally
depleted in the dinucleotide CpG
(its occurrence is only 21\% of
what would be expected under
randomness), the genome is
punctuated by regions with a high
frequency of CpG's relative to the
bulk genome.
This DNA characteristic is related to
the CpG methylation
\cite{international2001,gardiner1987}.
We may conclude that the clustering
of $DD_3$ has biological relevance.

It is worth noting that if one
considers all k-means clusterings
into 2 to 40 clusters, the second
best silhouette coefficient is
attained for 26 clusters, which
also corresponds to the point where
the CH index has a large increase
and the C-index is very small
($CH= 436$, $S=0.61$
and $C= 0.0046$).
In this partition with 26 clusters,
over half of the clusters are formed
by pairs of words that are reversed
complements of each other, i.e.,
obtained by reversing the order
of the word's symbols and
interchanging A-T and C-G.
The similarity between lag patterns of
reversed complements is a well-known
feature described in the literature,
see e.g. \cite{tavares2015}.

\subsection{Clustering words of
            length 5}

\begin{figure}[ht]
\centering
\includegraphics[width=0.8\textwidth]
  {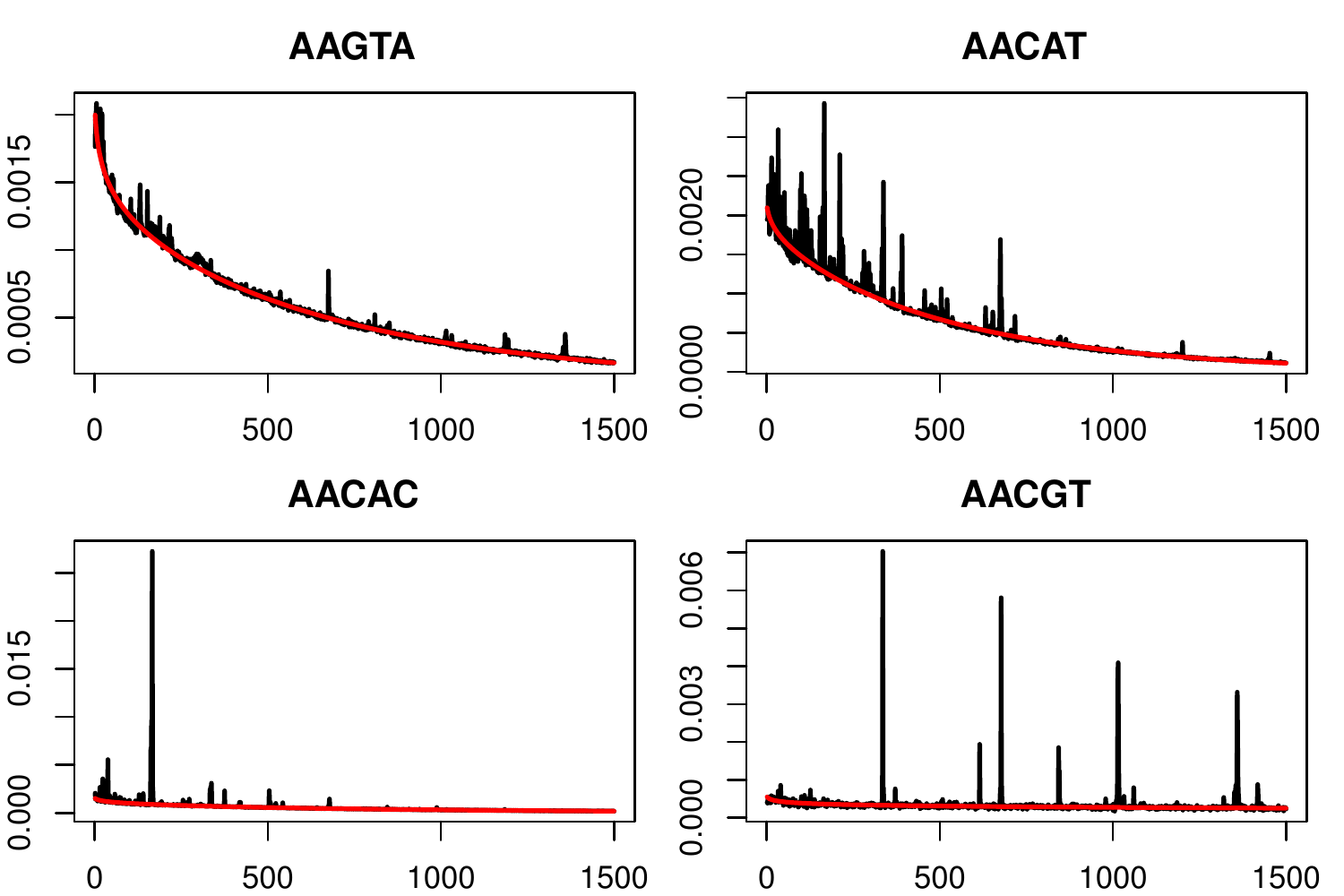}
\vskip-0.3cm
\caption{Lag distributions of some
  words of length $k=5$, with the
  corresponding baselines indicated
	in red.}
\label{fig:plotsk5}
\end{figure}

Also the lag distributions of $DD_5$
contain quite distinct baselines
and peak structures.
Figure~\ref{fig:plotsk5} shows four
lag distributions, with their
corresponding estimated baselines.

\begin{figure}[ht]
\centering
\includegraphics[width=1.0\textwidth]
  {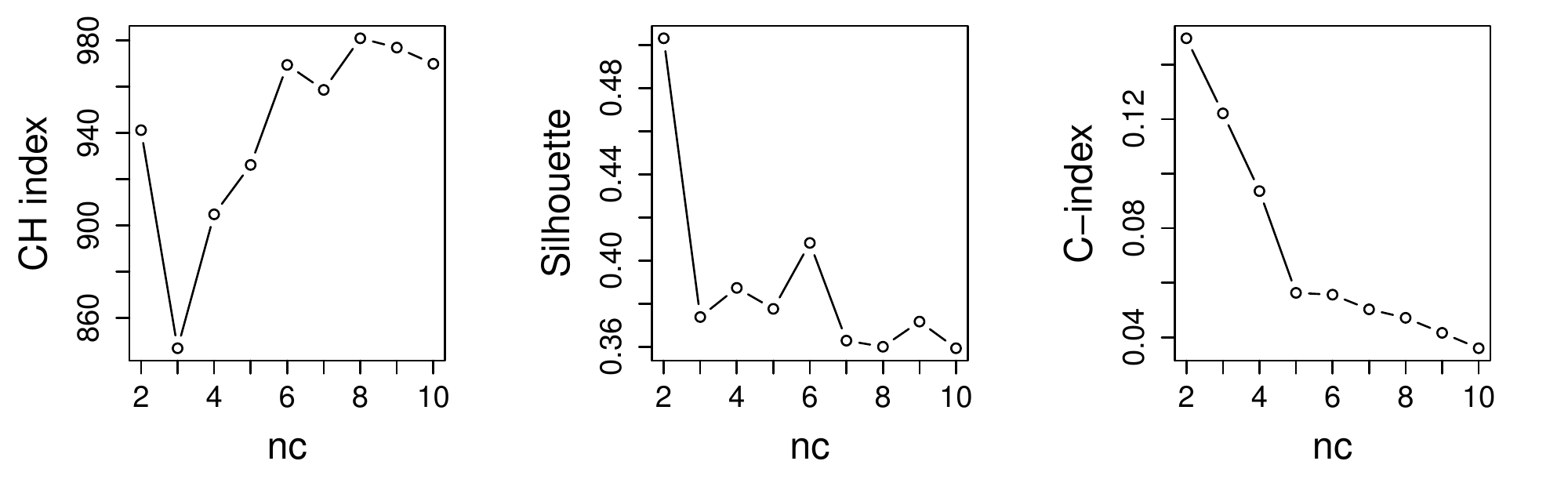}
\vskip-0.7cm
\caption{Validation indices for
clustering $DD_5$ by the number of
clusters $nc$: the Calinsky-Harabasz
index (left), average silhouette
width (center) and C-index (right).}
\label{fig:indexk5}
\end{figure}

Our procedure retains 3 principal
components
for the peaks and 2 components for
the baselines, so that each lag
distribution is converted into 5
scores.
Carrying out k-means clustering
for different numbers of clusters
yields the plots of validation
indices in Figure \ref{fig:indexk5}.
They do not all point to the same
choice, however.
The CH and S indices have local
maxima at 2 and 6 clusters, while
the C-index would support a choice
of 5 or more clusters.
It would appear that 2 or 6
clusters are appropriate.

When choosing 2 clusters we obtain
clusters with 278 and 746 members,
and when choosing 6 clusters they
have sizes 19, 92, 166, 141, 367
and 239.

We verified that both these
partitions are very stable. For this
we again drew 200 bootstrap samples,
and partitioned each of them
followed by computing the Jaccard
similarity coefficient of the
original clusters.
In the case of 2 clusters the
average Jaccard (stability) indices
were 0.94 and 0.97.
In the case of 6 clusters they were
0.84, 0.91, 0.93, 0.92, 0.93 and
0.93.
Since we aim to decompose the $DD_5$
dataset of 1024 distributions into
smaller groups with similar patterns,
we will focus on the solution with
6 clusters from here onward.

\begin{figure*}[htb]
\centering
\includegraphics[width=\textwidth]
  {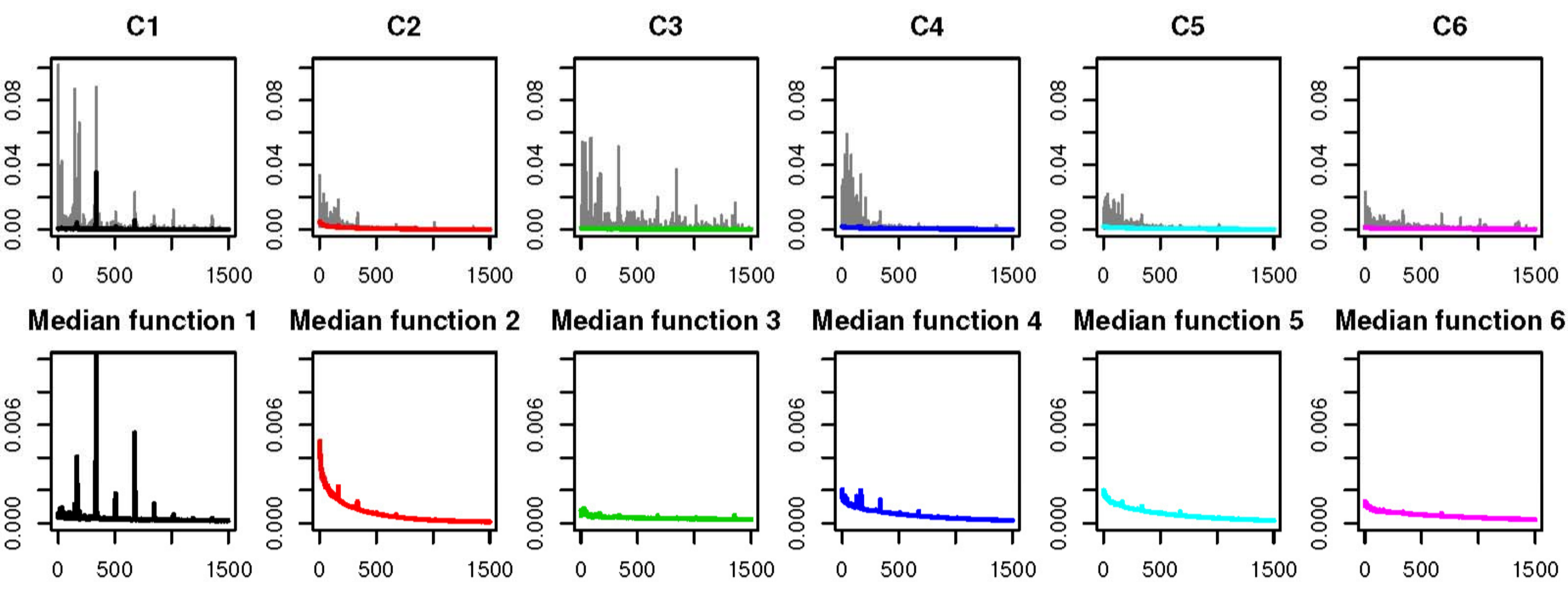}
\vskip-0.3cm
\caption{Clustering of $DD_5$ in six
clusters. In each cluster the lag
distributions are shown in grey, and
the cluster's median function is in
color (top). The median functions
are also shown with a scaled vertical
axis (bottom).}
\label{fig:clustk5}
\end{figure*}

The 6-cluster partition consists of
two large clusters
($|C_5|=367$ and $|C_6|= 239$),
three middle-sized clusters
($|C_2|= 92$, $|C_3|=166$ and
$|C_4|= 141$), and the much smaller
cluster $C_3$ with only 19 elements.
Figure \ref{fig:clustk5} shows the
lag distributions of each cluster.
As a graphical summary we also
consider the {\it median function}
of each cluster, which in each
domain point (lag) equals the
median of the cluster's function
values in that point.

We see the most pronounced peaks in
the clusters $C_1$, $C_3$ and $C_4$.
Those in the small cluster $C_1$
are the strongest.
Several of them occur in the same
location for most of the cluster
members, which explains why they
remain visible in the median function.
The words in $C_1$ are listed in
Table \ref{tab:k5_C3}.

\begin{table}[htbp]
\centering
\caption{List of words in cluster
$C_1$ of the partition of $DD_5$
in six clusters.}
\vskip0.3cm
\renewcommand{\arraystretch}{0.9}
\setlength\tabcolsep{3.5pt}
\normalsize
\begin{tabular}{llllll} 
\toprule
  AAACG & AACGG & ACGGG &
	AGCGC & CGAGA & CGCTT \\
	CGGGA & CGTTC & CGTTG &
	CTTCG & GAGGC & GCCTC \\
	GCGCT & GCGTT & TCGTA &
	TCGTT & TCTCG & TTCGT \\
	TTTCG\\
\bottomrule
\end{tabular}
\label{tab:k5_C3}
\end{table}

The distributions in $C_4$ have most
of their peaks before lag 500, with
little going on after that.
Cluster $C_3$ is quite different, as
strong peaks occur over the whole
domain.
The distributions in clusters $C_2$,
$C_5$ and $C_6$ have rather small
peaks, so few major irregularities.
Their main difference is in the
baselines: those of $C_2$ have a
high rate $\lambda$, whereas the
baselines of $C_6$ are much flatter.

We also explore the composition of
the words in each cluster, by
computing the percentage of words
that contain a given dinucleotide
or trinucleotide.
Clusters $C_1$, $C_2$ and $C_3$
stand out in this respect.
Cluster $C_2$ contains the largest
proportion of words with the
dinucleotides AA (47\%) and TT
(49\%), which is also reflected
in the high frequency of AAA and
TTT (25\% and 26\%, respectively).
The clusters $C_1$ and $C_3$ have
a lot of words containing the
dinucleotide CG (89\% and 98\%).
This is very different from the
other clusters:
only 9\% of the words in $C_2$
contain CG, in $C_4$ this is 11\%,
in $C_5$ only 1\%, and in
$C_6$ 16\%.
Even though both $C_1$ and $C_3$
have many CG dinucleotides, these
occur in different trinucleotides:
$C_1$ has many words
containing CGT and TCG (both 32\%),
whereas in $C_3$ many words contain
CGA (27\%) and ACG (23\%).

\section{Summary and conclusions}

In this work we have proposed a
methodology for decomposing the lag
distribution of a genomic word into
the sum of a baseline distribution
(the `trend') and a peak function.
The baseline component is estimated
by robustly fitting a parametric
function to the data distribution,
in which the residuals are made
homoskedastic and the robustness to
outliers is essential.
The peak function is then obtained
by comparing the absolute frequency
at each lag to a quantile of a
Poisson distribution.

When analyzing a dataset consisting
of many genomic words we can apply
principal component analysis to the
set of baselines and the set of
peak functions, which greatly
reduces the dimensionality.
This lower-dimensional data set
has uncorrelated scores and retains
much of the original information,
such as that in the Euclidean
distances.
This allows us to carry out k-means
clustering,
in which we have the choice whether
to use only the baseline information,
only the peak information, or both.
The performance of this approach was
evaluated by a simulation study,
which concluded that in situations
where both distinct baselines as
well as distinct peak functions
occur, the clustering procedure
using the combined information
performs very well.

This procedure was applied to the
data set $DD_3$ of all genomic
words of 3 symbols in human DNA,
as well as the set $DD_5$ of all
words of length 5.
This resulted in clusters of words
with specific distribution
patterns. By looking at the
composition of the words in each
cluster we found connections with
the frequency of certain
trinucleotides and dinucleotides,
such as CG which plays a particular
biological role.

Topics for further research are the
analysis of longer words, and the
application of other statistical
methods (such as classification)
on genomic data after applying the
decomposition technique developed
here.


\vskip0.8cm

\noindent{\large{\bf Acknowledgements}}

This work was partially supported by the Portuguese
Foundation for Science and Technology (FCT), Center
for Research \& Development in Mathematics and
Applications (CIDMA) and Institute of Biomedicine
(iBiMED), within projects UID/MAT/04106/2013 and
UID/BIM/04501/ 2013.
A. Tavares acknowledges the PhD grant
PD/BD/105729/ 2014 from the FCT.
The research of P. Brito was financed by the ERDF
- European Regional Development Fund through the
Operational Programme for Competitiveness and
Internationalization - COMPETE 2020 Programme
within project POCI-01-0145-FEDER-006961, and by
the FCT as part of project UID/EEA/50014/2013.
The research of J. \mbox{Raymaekers} and P. J.
Rousseeuw
was supported by projects of Internal Funds
KU Leuven.



\end{document}